\documentstyle[12pt]{article}
\newcommand{\be}{\begin{equation}}
\newcommand{\ee}{\end{equation}}
\newcommand{\ba}{\begin{eqnarray}}
\newcommand{\ea}{\end{eqnarray}}
\newcommand{\bb}{\begin{thebibliography}}
\newcommand{\eb}{\end{thebibliography}}
\newcommand{\ci}[1]{\cite{#1}}
\newcommand{\bi}[1]{\bibitem{#1}}

\begin {document}
\begin{center}
{\Large \bf ON THE OSCILLATIONS OF THE TENSOR SPIN
STRUCTURE FUNCTION}\\[1cm]
{A.V.EFREMOV, O.V.TERYAEV\\[0.3cm]
{\it Joint Institute for Nuclear Research} \\
{\it 141980, Dubna, Russia}}\\[1.2cm]
\end{center}



The tensor polarization is known to be the specific property of the
particles with spin larger than ${1\over 2}$. The deuteron is one of the
most fundamental spin-$1$ particles, and the effects of its tensor
polarization are intensively studied at low and intermediate energies. Such
effects should also be manifested for deep inelastic scattering (DIS) off
deuteron target, resulting in the new structure functions
\ci{FrStr,ET82,Jaf}. The DIS off longitudinally polarized deuterons have
recently been studied  by the Spin Muon Collaboration in order to extract
the neutron spin structure function $g_1^n$ \ci{SMC}.  The longitudinally
polarized deuteron target {\it automatically} receives the tensor
polarization as well, except the very special case when the probability of
zero spin projection is just $1/3$. It allows, in principle, the
tensor polarization effects to be studied by using the same data.

In this report, the simple description of the tensor polarization
in DIS is presented. We also remind and generalize a rather
old result \ci{ET82}; namely, the quark contribution to the tensor spin
structure function should manifest the oscillating behavior. Its
experimental study would allow one to discriminate between the deuteron
components with different spins.

The inclusive differential cross section for the DIS off spin-1 target
has the form:
\be
\sigma = p_+ \sigma_+ + p_- \sigma_- + p_0 \sigma_0,
\ee
where $p$'s are the probabilities of the corresponding projections and
$\sigma$'s are the cross sections for pure states. Usually, one uses
instead of $p$'s the (vector) polarization
\be
P=p_+ - p_-
\ee
and tensor polarization (alignment)
\be
T=p_+ + p_- - 2 p_0 = 1 - 3 p_0.
\ee

Using (2), (3) one can rewrite (1) in the form
\be
\sigma=\bar \sigma (1 + PA + {1\over 2} T A_T)
\ee
with the (vector) asymmetry
\be
A={\sigma_+ - \sigma_- \over {2 \bar \sigma}},
\ee
and the tensor asymmetry
\be
A_T={\sigma_+ + \sigma_- - 2 \sigma_0 \over {3 \bar \sigma}}.
\ee
The expression for the spin-averaged cross section is obvious
\be
\bar \sigma = {1 \over 3} (\sigma_+ + \sigma_- + \sigma_0).
\ee

The usually measured asymmetry for the target with polarization
"up" (parallel to the beam direction) and "down" is
\be
A^{exp} = {\sigma^{up} - \sigma^{down} \over {P(\sigma^{up} +
\sigma^{down})}}= {A\over {1+{1\over 2} T A_T}},
\ee
i.e. has a correction due to tensor asymmetry. The latter, however,
could be measured independently by using a nonpolarized target.
\be
A_T={\sigma^{up} + \sigma^{down} - 2 \bar \sigma \over {T \bar \sigma}}.
\ee

In the approximation of noninteracting proton and neutron, one easily gets
\ba
\sigma_+^d = \sigma_+^p + \sigma_+^n; \\
\sigma_-^d = \sigma_-^p + \sigma_-^n;  \\
\sigma_0^d = {1\over 2}(\sigma_+^p + \sigma_-^n)+
{1\over 2}(\sigma_+^n + \sigma_-^p)=\bar \sigma^p + \bar \sigma^n.
\ea
As a result, one has
\ba
\bar \sigma^d = \bar \sigma^p + \bar \sigma^n; \\
A^d = A^p {\bar \sigma^p \over {\bar \sigma^d}} +
A^n {\bar \sigma^n \over {\bar \sigma^d}};\\
A_T^d \equiv 0,
\ea
where $\bar \sigma^{p,n}={1\over 2} (\sigma_+^{p,n}+\sigma_-^{p,n})$ and
$A^{p,n}=(\sigma_+^{p,n} - \sigma_-^{p,n}) / 2 \bar \sigma^{p,n}$.
This means that the tensor asymmetry plays a very important role: it
measures the effect of deuteron boundness.

Note that the lepton beam polarization is inessential here. In fact, the
correlation between tensor and vector polarizations is related to the
antisymmetric part of the density matrix whose hermiticity results in a pure
imaginary factor. It should be compensated by the imaginary phase of the
scattering amplitude, absent in DIS (the only relevant momentum is spacelike).

The description of the tensor spin structure function in the parton model
should naively be very different in the case of the partons with different
spins -- quarks and gluons.  Here the sum rules for the
first two moments are proposed, which are just the consequence of
this difference.  They also discriminate between
hadronic components of the deuteron with a different spin.  Their validity
and violation should provide the important information about the nucleon
and deuteron spin structure.

It should be mentioned that the sum rules of interest were proposed
already in 1982 \ci{ET82}. Although the high-$p_T$ vector meson in the
{\it final} state was considered, the quark contribution to the
tensor spin structure function was defined just for the initial state
case. Let us briefly recall this definition. The quark
contribution to the part of the cross section, proportional to $T$ can be
expressed as
\be
\label{F}
\sigma_q=\int d^4 z tr[E_{\mu ^2}\gamma^\nu]
\langle P, S|\bar \psi (0) \gamma _\nu \psi(z) |P, S \rangle _{\mu ^2}
\ee
Here $E$ is the short-distance part, ${\mu ^2}$ being its IR
regularization parameter, the same as the UV one for the matrix element. The
Taylor expansion of the latter results in the obvious parton formula
\be
\sigma_q=\int^1_0 dx tr[\hat P E_{\mu ^2} (xP)] C^T_{\mu ^2} (x) s^{zz}
\ee
with the moments of quark tensor spin structure
distribution related to the matrix elements of the local composite
operators
\be
\langle P, S|\bar \psi (0) \gamma ^\nu D^{\nu_1}...D^{\nu_n} \psi(0)
|P, S \rangle _{\mu ^2}=i^{-n} M^2 S^{\nu \nu_1} P^{\nu_2}...P{\nu_n}
\int_0^1 C^T_q (x) x^n dx.
\ee
Here $S^{\mu \nu}$ is the traceless symmetric tensor providing the
covariant description of the vector meson alignment.
In hard processes, the single component $S^{\mu \nu}=s^{zz} P^{\mu} P^{\nu}
/M^2$ dominates, where $s^{ij}$ is the Cartesian spin-tensor in
the target rest frame, the latter being directly related to $T$.
This is quite analogous to the dominance of longitudinal vector polarization
and kinematical suppression of the transverse one. Only this dominant
contribution was considered in the paper \ci{ET82}.
The full analysis \ci{Jaf}, however, also leads to the identification of
the dominant structure function.

The zero sum rule for each quark flavor $i$ follows immediately, just
because the matrix element for $n=0$ vanishes:
\be
\label{SR}
\int_0^1 C^T_i (x) dx = 0.
\ee
This "naive" derivation is quite analogous to that of
the Burkhardt-Cottingham sum rule in QCD \ci{AhRo,ET82a}. The problem of
its possible violation is still discussed \ci{JafJi}. However, there are
solid arguments \ci{IoLi} against such a violation in the scaling region.

Note that the $n=1$ operator is just the quark contribution to the
energy-momentum tensor. Taking into account the contributions
of all flavors \ci{dis} and gluons, one should get the
$S^{\mu \nu}-$and ${\mu ^2}-$independent matrix element
fixed by the energy-momentum conservation:
\be
\sum_{q,g} \langle P, S|T_i^{\mu \nu}|P, S \rangle _{\mu ^2}=
2 P^{\mu} P^{\nu}.
\ee
However, quark and gluon contributions may, in principle, depend on
$S^{\mu \nu}$:
\ba
\sum_{q} \langle P, S|T_i^{\mu \nu}|P, S \rangle _{\mu ^2}=
2 P^{\mu} P^{\nu} (1-\delta(\mu^2)) +2 M^2 S^{\mu \nu} \delta_1(\mu^2)\\
\langle P, S|T_g^{\mu \nu}|P, S \rangle _{\mu ^2}=
2 P^{\mu} P^{\nu} \delta(\mu^2) - 2 M^2 S^{\mu \nu} \delta_1(\mu^2)
\ea
This natural parametrization results in the gluonic correction to the
$n=1$ zero sum rule:
\be
\sum_q \int_0^1 C^T_i (x) x dx = \delta_1(\mu^2).
\ee

The gluons contribute to the $n=0$ sum rule as well. There is an
additional contribution to the cross section, equal to the convolution
of the gluon coefficient function with the gluon tensor distribution.
The latter may have a non-zero first moment contrary to the quark one
\be
\langle P, S| O_g^{\nu \nu_1} |P, S \rangle _{\mu ^2}=M^2 S^{\nu \nu_1}
\int_0^1 C^T_g (x) dx.
\ee
Here $O_g^{\mu \nu}$ is the (renormalized) local gluonic operator.
It may be  constructed
either from the gauge-invariant field strength or the gluon field itself in
the "physical" axial gauge. This contribution, however, is suppressed
by $\alpha_s$ entering in the coefficient function. The gluon
contribution to the deuteron tensor structure function, associated with
the box diagram, was calculated a few years ago \ci{JaMa}. It is similar
to the gluon contribution to the linear polarized photon structure
function. The authors therefore claimed that the deuteron should be aligned
perpendicular to the beam. This statement naively contradicts the
kinematical dominance of the longitudinal alignment mentioned above.
However, the tensor polarizations in the mutually orthogonal directions
are not independent because the tensor $S^{\mu \nu}$ is traceless. The sum
of $\rho_{00}^i$, the zero spin projection probabilities, over three
orthogonal directions $i$ is equal to unity. If the target is aligned
along the beam direction, the rotational symmetry leads to \ci{ET82}:
\be
\rho_{00}^L+2 \rho_{00}^T = 1.
\ee
The transverse alignment is absent ($\rho_{00}^T = 1/3$) if and only if
the longitudinal one is also absent.

Note that zero sum appeared to be valid, provided the unobserved long-range
singularity is taken into account \ci{JaMa}. Such a possibility was also
considered in the case of the Burkhardt-Cottingham sum rule \ci{AhRo,JafJi}.

The sum rule (\ref{SR}) means, of course, that $C_T$ should change sign
somewhere.
If $\delta_1$ is numerically small, the oscillations of the singlet
tensor distributions are even more dramatic. It crosses zero at least
at two  points. It is interesting that the model calculations of the
tensor distribution \ci{Jaf,Mank} really manifest an oscillating behavior.
The parton model analysis \ci{ClKu} shows that its violation is caused
by the deuteron quadrupole structure.

It is possible to describe such a behavior by considering a more
conservative approach to the deuteron. Formula (\ref{F}) is still valid if
the quarks and gluons are replaced by the hadrons (nucleons and mesons).
This is a straightforward generalization of the operator product
expansions using the basis consisting of hadronic local operators
\ci{Kap,Petr}. One may conclude that the zero sum rule is valid,
as far as nucleonic operators (analogous in this sense to the quark ones)
are considered. It is also valid
for the (pseudo)scalar operators constructed from pion fields. However,
it is obviously violated by the operators constructed
from vector meson fields "substituting" the gluon ones in this approach.

It is very interesting to study the zero sum rule experimentally.
The tensor structure function can be measured, as it has been mentioned
above, by the Spin Muon Collaboration. However, as it is probably
numerically small in comparison with $g_1$, one may expect to obtain some
restrictions from above only. Nevertheless, even such a result would be
important as a check of the validity of free nucleon approximation (it is
used in order to extract the neutron spin structure function).  Moreover,
one should take into account the tensor asymmetry in order to extract
the vector one in the self-consistent way.  One cannot exclude the
enhancement of the tensor structure function in some kinematical region,
making possible its measurement by SMC. If it should happen  at low $x$
region, the first moment of $g_1^n$, entering in the Bjorken sum rule, may
be affected significantly.

To study the tensor spin structure more statistics is required.
This probably should be done by the HERMES collaboration at HERA \ci{HERA}
and, possibly, by the European Electron Facility \ci{Kum}. It seems
possible to do this also at CEBAF, simultaneously with the
already proposed \ci{CEBAF} study of generalized Gerasimov--Drell--Hearn
sum rule.

We conclude that the experimental study of the first two moments of
the deuteron tensor spin structure function can provide some information
about its constituents with different spins.

It is a pleasure to thank M. Anselmino, L. Kaptari, K. Kazakov and
E. Leader for useful discussions. 

This work was supported in part by the Russian Foundation for
Fundamental Researches Grant $N^0$ 93-02-3811.

\end{document}